\documentclass[prd,aps,showpacs,amsmath,amssymb,nofootinbib]{revtex4}
\usepackage{graphicx}% Include figure files
\usepackage{dcolumn}% Align table columns on decimal point
\usepackage{bm}% bold math

\begin{document}

%%%%
%    Greek Letters
%

\let\a=\alpha      \let\b=\beta       \let\c=\chi        \let\d=\delta
\let\e=\varepsilon \let\f=\varphi     \let\g=\gamma      \let\h=\eta
\let\k=\kappa      \let\l=\lambda     \let\m=\mu
\let\o=\omega      \let\r=\varrho     \let\s=\sigma
\let\t=\tau        \let\th=\vartheta  \let\y=\upsilon    \let\x=\xi
\let\z=\zeta       \let\io=\iota      \let\vp=\varpi     \let\ro=\rho
\let\ph=\phi       \let\ep=\epsilon   \let\te=\theta
\let\n=\nu
\let\D=\Delta   \let\F=\Phi    \let\G=\Gamma  \let\L=\Lambda
\let\O=\Omega   \let\P=\Pi     \let\Ps=\Psi   \let\Si=\Sigma
\let\Th=\Theta  \let\X=\Xi     \let\Y=\Upsilon

%
%%%

%%%
%    Calligraphic letters
%

\def\cA{{\cal A}}                \def\cB{{\cal B}}
\def\cC{{\cal C}}                \def\cD{{\cal D}}
\def\cE{{\cal E}}                \def\cF{{\cal F}}
\def\cG{{\cal G}}                \def\cH{{\cal H}}
\def\cI{{\cal I}}                \def\cJ{{\cal J}}
\def\cK{{\cal K}}                \def\cL{{\cal L}}
\def\cM{{\cal M}}                \def\cN{{\cal N}}
\def\cO{{\cal O}}                \def\cP{{\cal P}}
\def\cQ{{\cal Q}}                \def\cR{{\cal R}}
\def\cS{{\cal S}}                \def\cT{{\cal T}}
\def\cU{{\cal U}}                \def\cV{{\cal V}}
\def\cW{{\cal W}}                \def\cX{{\cal X}}
\def\cY{{\cal Y}}                \def\cZ{{\cal Z}}

\def\dbd{{$0\nu 2\beta\,$}}
%
%%%%

\newcommand{\Ns}{N\hspace{-4.7mm}\not\hspace{2.7mm}}
\newcommand{\qs}{q\hspace{-3.7mm}\not\hspace{3.4mm}}
\newcommand{\ps}{p\hspace{-3.3mm}\not\hspace{1.2mm}}
\newcommand{\ks}{k\hspace{-3.3mm}\not\hspace{1.2mm}}
\newcommand{\des}{\partial\hspace{-4.mm}\not\hspace{2.5mm}}
\newcommand{\desco}{D\hspace{-4mm}\not\hspace{2mm}}

%%%%
%\usepackage[dvips]{graphicx}
%\usepackage{yfonts}
\newcommand{\beq}{\begin{eqnarray}}
\newcommand{\eeq}{\end{eqnarray}}
\renewcommand\d{\partial}
\newcommand{\tr}{\mathop{\mathrm{tr}}}
%%%%%%%%%%%%%%%%%%%%%%%%%%%%%%%%%%%%%555
%\begin{document}
 \title{\boldmath Asymmetries and observables for $\Lambda_b\rightarrow \Lambda \ell^+\ell^-$}

\author{Girish Kumar$^{a,b}$ and Namit Mahajan$^a$
}
\email{girishk@prl.res.in, nmahajan@prl.res.in}
\affiliation{
 {$^a$Theoretical Physics Division, Physical Research Laboratory, Navrangpura, Ahmedabad
380 009, India }\\
{$^b$ Department of Physics, Indian Institute of Technology, Gandhinagar, 382 424, India}
}

%%%%%%%%%%%%%%%%%%%%%%%%%%%%%%%%%%%%%%%%%%
\begin{abstract}
The semi-leptonic baryonic $b\to s$ decay, $\Lambda_b\rightarrow \Lambda \ell^+\ell^-$, has been studied and 
new angular observables and asymmetries have been proposed which can test the presence of new physics beyond the standard model.
 %
 %In our analysis, we also consider
 %helicity flipped operators, namely, $O_7^{'}$, $O_9^{'}$ and $O_{10}^{'}$ to obtain the relations. 
 %We also construct a new observable of interest, call it 
 %$\mathcal{O}_T^{L,R}$, which in SM limit has positions of zero exactly at zero of forward-backward asymmetry
 %zero but in presence of NP contributions will show different 
 %behavior than that of zero of AFB.  We point out that the precise measurements of these zeroes in near future
 %would provide crucial test of SM and 
 %would be useful in distinguishing between different possible new physics contribution in Wilson coefficients. 
%
\end{abstract}

\maketitle

\section{Introduction}
Rare decays of the b-quark offer a unique possibility to study the weak interactions operating at the fundamental level
governing the decays in conjunction with the strong forces responsible for keeping the constituents bound in various
colourless hadronic states. The large mass of the b-quark compared to the typical QCD scale ensures that the perturbative
hard part can be factorized from the long distance hadronic dynamics. It is in fact the long distance hadronic dynamics
which is at the heart of the problem of obtaining accurate and reliable results. After factorizing the short and long distance
pieces, the hadronic matrix elements (to be defined below in detail for the relevant case) are expressed in terms
of the form factors which carry the non-trivial $q^2$ dependence, where $q$ generically denotes the momentum transfer
for the process in question (see \cite{Buchalla:1995vs} for a review). 
Being of non-perturbative origin, these form factors need to be calculated via methods like
lattice methods or QCD sum rules. On the perturbative side, one can hope to compute the short distance pieces
(called Wilson coefficients) to higher order and ensure better accuracy and stability of the results. 

Semi-leptonic decays mediated by the quark level transition $b\to s \ell^+\ell^-$ offer cleaner probes compared to
non-leptonic exclusive hadronic decays. In the latter case, theoretical calculations are more difficult in general 
and they are also marred with issues related to the QCD effects, both perturbative and non-perturbative, in a bigger
way. Semi-leptonic decays on the other hand are somewhat easier at the theoretical level as the leptonic sub-system
factorizes as far as the QCD effects between the final state sub-systems go. Further, since LHCb hints at
deviations from the standard model (SM) predictions in observables related to $B\to K^{(*)}\mu^+\mu^-$ channels
(see \cite{Aaij:2013qta}  for anomalies in $K^*$ channel and \cite{Aaij:2014ora} for hints of lepton universality violation
in $K$  channel),
which proceed at the quark level by the same $b\to s$ semi-leptonic decay, it is of utmost importance
to study any other such semi-leptonic decay modes to clarify the situation and pin point the source of
these deviations. Since the hadronic effects bring along large uncertainties, the above said hints can not be
conclusively taken as evidence for new physics, which is part of the short distance structure. 
However, if similar pattern emerges for decays with different hadronic particles but governed by the same
$b\to s\ell^+\ell^-$ quark level transition, then that would amount to unambiguous signal for physics beyond SM. The baryonic
decay $\Lambda_b\rightarrow \Lambda \ell^+\ell^-$ satisfies all these requirements and therefore it is useful 
to study it in detail. This decay has been studied theoretically in the past \cite{Huang:1998ek} but the emphasis has been
somewhat different from that in this paper. This decay mode, like $B\to K^{*}\ell^+\ell^-$ has many angular 
observables to offer as probes. This fact was utilized to some extent in \cite{Boer:2014kda}. Here we take it further and also
construct some new angular observables which are theoretically clean and do not depend sensitively on
the hadronic form factors.
On the experimental side, this decay was observed at the Tevatron \cite{Aaltonen:2011qs}. Recently, LHCb has measured
the branching fraction along with some
angular coefficients \cite{Aaij:2013mna}, \cite{Aaij:2015xza}. The errors are still quite large but one hopes to have better results in near future.

%%%%%%%%%%%%%%%%%%%%%%%%%%%%%%%%%%%%%%%%%%%%%%%%%%5
\section{Effective Hamiltonian and the decay $\Lambda_b\rightarrow \Lambda \ell^+\ell^-$}
 The framework to study such rare decays %where QCD corrections can give sizeable contributions 
 is that
 of effective Hamiltonian which 
 is obtained after integrating out the heavy degrees of freedom. 
 The rare decay $\Lambda_b\rightarrow \Lambda \ell^+\ell^-$ is governed by effective Hamiltonian% for $ b\rightarrow s $ transitions written as  
\begin{equation}\label{eq:eff H}
 \mathcal{H}_{eff} = -\frac{4 G_F}{\sqrt{2}} V_{ts}^{*} V_{tb} \sum_i C_i (\mu)O_i  + \mbox{h.c.}
\end{equation}
where contribution of the term $\propto \frac{V_{ub} V_{us}^{*}}{V_{tb} V_{ts}^{*}}$ is neglected. 
$O_i$ are the effective  
local operators and $C_i(\mu)$ are called Wilson coefficients evaluated at scale $\mu$. 
The factorization scale $\mu$ distinguishes between
short distance physics (above scale $\mu$) and long distance physics (below scale $\mu$).
Wilson coefficients encode information about heavy degrees of freedom which have been integrated out while
matrix elements of local operators $O_i$
dictate the low energy dynamics (for a review see \cite{Buchalla:1995vs}). 

The operators contributing significantly to the process $b\to s\ell^+\ell^-$ in SM are 
%semi-leptonic vector operator 
%$O_9$, axial vector operator 
%$O_{10}$ and magnetic photon penguin operator $O_7$. Their explicit form is given by
\begin{equation}\label{eq:operators}
 \begin{split}
 O_7 &= \frac{e}{16 \pi^2} m_b (\bar{s}_\alpha \sigma_{\mu\nu} R b_\alpha) F^{\mu\nu}, \\
%  O_8 = \frac{g_s}{16 \pi^2} m_b(\bar{s}_\alpha T^a_{\alpha\beta}\sigma_{\mu\nu} R b_\beta)G^{a\mu\nu},\\
 O_9 &= \frac{e^2}{16 \pi^2}(\bar{s}_\alpha\gamma^\mu L b_\alpha)(\bar{l} \gamma_\mu l),\\
 O_{10} &= \frac{e^2}{16 \pi^2}(\bar{s}_\alpha\gamma^\mu L b_\alpha)(\bar{l} \gamma_\mu\gamma_5 l),
 \end{split}
\end{equation}
where, $\alpha,\,\,\beta \,$ are the color indices,  $L,R = \frac{1}{2}(1\mp\gamma_5)$ represent chiral projections, 
$T^a$ are the  SU(3) color charges and 
$m_b$ is the b-quark mass. All the information about short distance physics 
and possible new physics effects is contained in the Wilson coefficients which are computed in perturbation theory 
as a series in $\alpha_s$ and evaluated at the scale $\mu$ using the renormalization group equations. Beyond SM, there could
be new operators with flipped helicity structure or different tensorial structure like scalar-pseudoscalar operators.
%The ultimate goal is to establish if any new physics beyond SM is present, and if there, what is its flavour structure.

At first sight the decay $\Lambda_b\rightarrow \Lambda \ell^+\ell^-$ may seem not to be too useful owing to larger
uncertainties in the transition form factors involved, when compared to the mesonic counterpart $B\rightarrow K^* \ell^+\ell^-$.
However, this decay offers a larger number of observables. For example, in contrast to $K^*\to K \pi$ decay in 
the mesonic counterpart which is parity conserving, $\Lambda\to N\pi$ is a parity violating decay 
and hence brings along the possibility of measuring forward-backward asymmetry in the hadronic system as well. This decay
has been studied theoretically, but the emphasis in most of those studies was mainly on the lepton forward-backward asymmetry
and/or lepton polarization asymmetry. Since the decay was observed at Tevatron, there has been some activity,
both on the form factors \cite{Detmold:2012vy} as well as exploiting the angular observables \cite{Boer:2014kda}. In the present
work, we extend the analysis of \cite{Boer:2014kda} and also propose new observables and asymmetries which are 
theoretically clean and can be used with the limited data expected in near future.

The four body differential decay $\Lambda_b (p) \rightarrow \Lambda (k) [\to N (k_1)\pi(k_2)] \ell^+(q_1)\ell^-(q_2)$
can be conveniently written in terms of the variables: invariant mass squared of the lepton system $q^2=(p-k)^2$, 
helicity angles $\theta_{\Lambda}$ and $\theta_{\ell}$ of the hadronic and leptonic sub-systems respectively,
and the azimuthal angle $\phi$ between the hadronic and leptonic planes. Taking into account the polarizations
of $\Lambda_b$ and $\Lambda$, there are a host of form factors that enter the calculations (see \cite{Boer:2014kda} for details).
The four body differential decay rate can be written as

\begin{equation}
 \frac{d^4\Gamma}{dq^2d\cos\theta_{\ell}d\cos\theta_{\Lambda}d\phi} = \frac{3}{8\pi}K(q^2,\cos\theta_{\ell},\cos\theta_{\Lambda},\phi)
\end{equation}
where
\begin{eqnarray}
 K(q^2,\cos\theta_{\ell},\cos\theta_{\Lambda},\phi) &=& K_{1ss}\sin^2\theta_{\ell} + K_{1cc}\cos^2\theta_{\ell} + 
 K_{1c}\cos\theta_{\ell} \nonumber \\
 &+& (K_{2ss}\sin^2\theta_{\ell} + K_{2cc}\cos^2\theta_{\ell} +  K_{2c}\cos\theta_{\ell}) \cos\theta_{\Lambda} \\
 &+& (K_{3sc}\sin\theta_{\ell}\cos\theta_{\ell}+ K_{3s}\sin\theta_{\ell}) \sin\theta_{\Lambda}\cos\phi \nonumber \\
 &+& (K_{4sc}\sin\theta_{\ell}\cos\theta_{\ell}+ K_{4s}\sin\theta_{\ell}) \sin\theta_{\Lambda}\sin\phi \nonumber
\end{eqnarray}
The angular coefficients $K_{..}$ depend only on the dilepton invariant mass, $q^2$, and carry the 
hadronic information. They are in turn expressed in terms of the transversity amplitudes. These transversity
amplitudes are written as combinations of Wilson coefficients and baryonic form factors. A typical form factor is
denoted as $H(s_{\Lambda_b},s_{\Lambda})$, where we have suppressed the indices $V,\,T,\, A$ signifying the type of
operator sandwiched between the external hadronic states but have explicitly shown the two spin projection vectors which
take values $\pm 1/2$.

       \begin{figure}[ht!]
   \vskip 0.32cm
   \hskip 1.35cm
   \hbox{\hspace{0.03cm}
   \hbox{\includegraphics[scale=0.30]{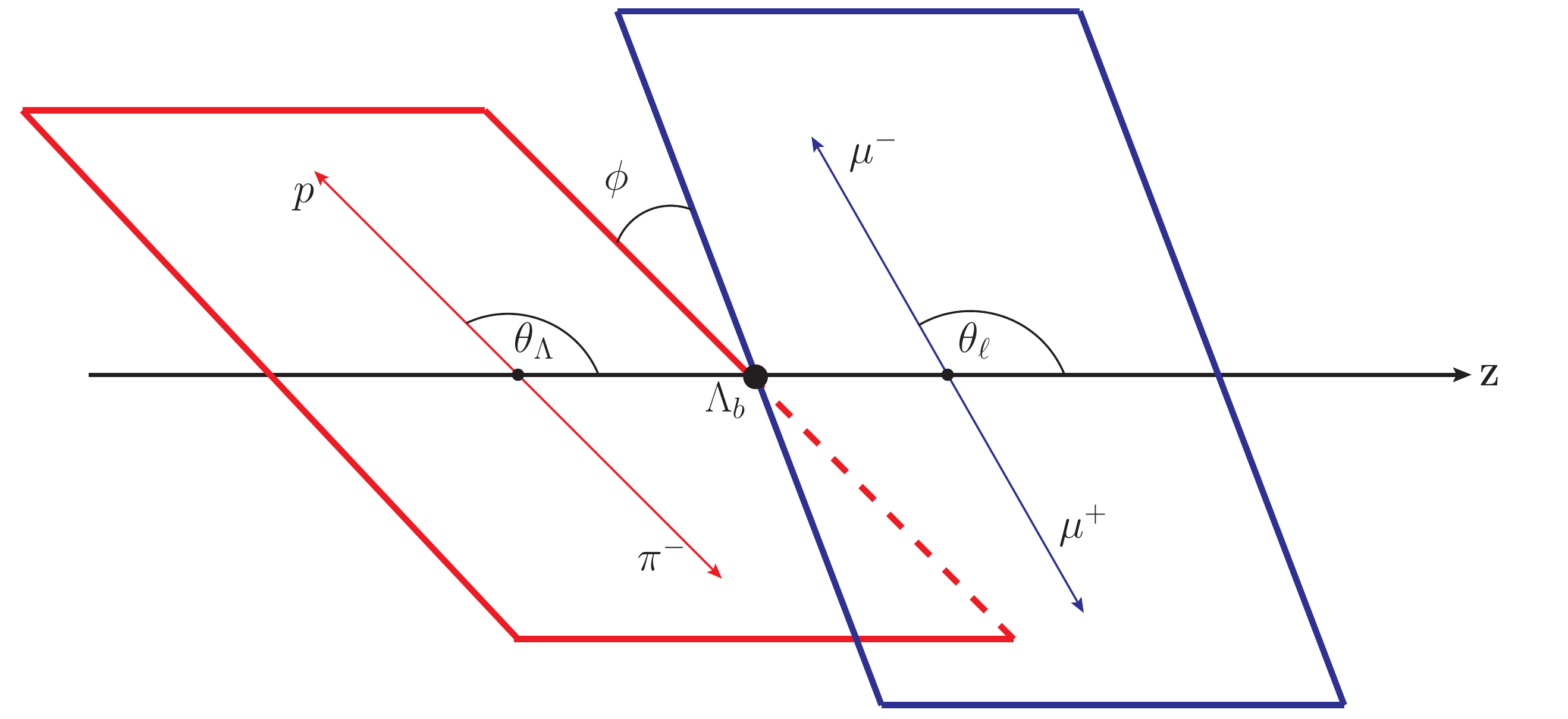}}
   }
  
   \caption{The schematic diagram of  the angular distribution of $\Lambda_b  \rightarrow \Lambda (\to N \pi) \ell^+\ell^-$ decay with the description of the angles $\theta_{\ell}$, $\theta_{\Lambda}$ and $\phi$.
    }
    \label{fig1}
       \end{figure}
 In SM, the transversity amplitudes read
 \begin{eqnarray}
  A_{\perp_1}^{L(R)} &=& \sqrt{2} N \left(C_{9,10}^{L(R)} H_{+}^V(-1/2, 1/2) - 
  \frac{2 m_b C_7 }{q^2} H_{+}^T (-1/2, 1/2)\right) \nonumber \\
A_{\parallel_1}^{L(R)} &=& -\sqrt{2} N \left(C_{9,10}^{L(R)} H_{+}^A(-1/2, 1/2) 
+ \frac{2 m_b C_7}{q^2} H_{+}^{T5} (-1/2, 1/2)\right) \nonumber \\
A_{\perp_0}^{L(R)} &=& \sqrt{2} N \left(C_{9,10}^{L(R)} H_{0}^V(1/2, 1/2) 
- \frac{2 m_b C_7 }{q^2} H_{0}^T (1/2, 1/2)\right)\\
A_{\parallel_0}^{L(R)} &=& -\sqrt{2} N \left(C_{9,10}^{L(R)} H_{0}^A(1/2, 1/2) 
+ \frac{2 m_b C_7}{q^2} H_{0}^{T5} (1/2, 1/2)\right)\nonumber
 \end{eqnarray}
 where $C_{9,10}^{L(R)} = (C_9 \mp C_{10})$ and 
 $N = G_F V_{tb} V_{ts}^{*} \alpha_e \sqrt{\frac{q^2 \sqrt{\lambda(m_{\Lambda_b}^2, m_{\Lambda}^2, q^2)}}
 {3\,\pi^5(2)^{11}\,m_{\Lambda_b}^3}}$ is the normalization factor.
 
In terms of these transversity amplitudes, the angular coefficients appearing in the fully differential
decay rate read
\begin{eqnarray}
 K_{1ss} = \frac{1}{4}\left[ |A_{\perp_1}^R|^2 + |A_{\parallel_1}^R|^2 + 2|A_{\perp_0}^R|^2 + 2|A_{\parallel_0}^R|^2 + 
 ( R \leftrightarrow L)\right]\nonumber \\
  K_{1cc} = \frac{1}{2}\left[ |A_{\perp_1}^R|^2 + |A_{\parallel_1}^R|^2 + ( R \leftrightarrow L)\right] \nonumber \\
  K_{1c} = -\mbox{Re}\left\{A_{\perp_1}^R A_{\parallel_1}^{*R} -( R \leftrightarrow L) \right\} \nonumber \\
    K_{2ss} = \frac{\alpha}{2}\mbox{Re}\left\{A_{\perp_1}^R A_{\parallel_1}^{*R}  +
    2A_{\perp_0}^R A_{\parallel_0}^{*R} + ( R \leftrightarrow L) \right\} \nonumber \\
  K_{2cc} = \alpha\mbox{Re}\left\{A_{\perp_1}^R A_{\parallel_1}^{*R}  + ( R \leftrightarrow L) \right\} \\
  K_{2c} = -\frac{\alpha}{2}\left[ |A_{\perp_1}^R|^2 + |A_{\parallel_1}^R|^2 - ( R \leftrightarrow L)\right] \nonumber \\
  K_{3sc} = \frac{\alpha}{\sqrt{2}}\mbox{Im}\left\{A_{\perp_1}^R A_{\perp_0}^{*R} -
  A_{\parallel_1}^R A_{\parallel_0}^{*R} + ( R \leftrightarrow L)\right\} \nonumber \\
  K_{3s} = \frac{\alpha}{\sqrt{2}}\mbox{Im}\left\{A_{\perp_1}^R A_{\parallel_0}^{*R} - 
  A_{\parallel_1}^R A_{\perp_0}^{*R} - ( R \leftrightarrow L)\right\} \nonumber \\
  K_{4sc} = \frac{\alpha}{\sqrt{2}}\mbox{Re}\left\{A_{\perp_1}^R A_{\parallel_0}^{*R} - 
  A_{\parallel_1}^R A_{\perp_0}^{*R} + ( R \leftrightarrow L)\right\} \nonumber \\
  K_{4s} = \frac{\alpha}{\sqrt{2}}\mbox{Re}\left\{A_{\perp_1}^R A_{\perp_0}^{*R} - 
  A_{\parallel_1}^R A_{\parallel_0}^{*R} - ( R \leftrightarrow L)\right\} \nonumber
\end{eqnarray}
where the parameter $\alpha$ is the parity violating parameter in the $\Lambda\to N\pi$ decay.

The task then is to experimentally determine these angular coefficients. In principle, once there is sufficient data,
a full angular fit would end up determining these coefficients (up to discrete ambiguities). One could
proceed by studying angular asymmetries allowing for the extraction of specific angular coefficients and/or
some combinations of those. In \cite{Boer:2014kda}, the authors considered the following observables which provide a handle on a select few 
angular coefficients:\\
(i) Decay rate as a function of $q^2$
\begin{equation}
 \frac{d\Gamma}{dq^2} = 2K_{1ss} + K_{1cc} \label{rate}
\end{equation}
(ii) Transverse (and therefore longitudinal) polarization fraction
\begin{equation}
 F_L = 1 - F_T = \frac{2K_{1ss} - K_{1cc}}{2K_{1ss} + K_{1cc}} \label{FL}
\end{equation}
(iii) Forward-backward asymmetries in the leptonic, hadronic and mixed sub-systems:
\begin{eqnarray}
A_{FB}^{\ell} &=& \frac{3}{2}\frac{K_{1c}}{2K_{1ss} + K_{1cc}} \nonumber \\
A_{FB}^{\Lambda} &=& \frac{1}{2}\frac{2K_{2ss} + K_{2cc}}{2K_{1ss}+K_{1cc}} \\
A_{FB}^{\ell,\Lambda} &=& \frac{3}{4}\frac{K_{2c}}{2K_{1ss}+K_{1cc}}\nonumber \label{Afb}
\end{eqnarray}
Analogous to the lepton forward-backward asymmetry in $B\rightarrow K^* \ell^+\ell^-$, $A_{FB}^{\ell}$ and 
$A_{FB}^{\ell,\Lambda}$ have a zero crossing, which essentially depends on the short distance parameters only 
(in the approximation when the form factor dependence more or less cancels) and its value is same as in the the mesonic
case, scaled by the $\Lambda_b$ mass instead of the B-meson mass.

\section{More asymmetries and new observables}
We extend the previous work by constructing asymmetries such that all the angular coefficients can be
extracted. To this end, we construct the following observables: \\
\begin{eqnarray}
 {Y_2} &=& \frac{\int_{0}^{2\pi} d\phi \left[\int_{0}^{1} -\int_{-1}^0\right] d\cos\theta_\Lambda 
 \left[\int_{-1}^{-1/2} -\int_{-1/2}^{0} - \int_{0}^{1/2} +\int_{1/2}^1\right] d\cos\theta_{\ell} 
 K(q^2,\theta_{\ell},\theta_\Lambda,\phi)}
 {\int_{0}^{2\pi} d\phi \int_{-1}^1 d\cos\theta_\Lambda \int_{-1}^{1} d\cos\theta_{\ell} 
 K(q^2,\theta_{\ell},\theta_\Lambda,\phi)} \label{Y2}\\ 
     &=& \frac{3}{8}\frac{K_{2cc} - K_{2ss}}{2K_{1ss}+K_{1cc}} \nonumber 
\end{eqnarray}

\begin{eqnarray}
 {Y_{3s}} &=&  \frac{\left[\int_{0}^{\pi/2} -\int_{\pi/2}^{\pi} - \int_{\pi}^{3\pi/2} + \int_{3\pi/2}^{2\pi}\right]
 d\phi \int_{-1}^{1} d\cos\theta_\Lambda 
 \int_{-1}^{1} d\cos\theta_{\ell} K(q^2,\theta_{\ell},\theta_\Lambda,\phi)}
 {\int_{0}^{2\pi} d\phi \int_{-1}^1 d\cos\theta_\Lambda \int_{-1}^{1} d\cos\theta_{\ell} 
 K(q^2,\theta_{\ell},\theta_\Lambda,\phi)} \label{Y3s}\\ 
     &=& \frac{3\pi}{8}\frac{K_{3s}}{2K_{1ss}+K_{1cc}} \nonumber 
\end{eqnarray}

\begin{eqnarray}
 {Y_{3sc}} &=&  \frac{\left[\int_{0}^{\pi/2} -\int_{\pi/2}^{\pi} - \int_{\pi}^{3\pi/2} + \int_{3\pi/2}^{2\pi}\right] 
 d\phi \int_{-1}^{1} d\cos\theta_\Lambda \left[\int_{0}^{1} -\int_{-1}^0\right] d\cos\theta_{\ell} 
 K(q^2,\theta_{\ell},\theta_\Lambda,\phi)} {\int_{0}^{2\pi} d\phi \int_{-1}^1 d\cos\theta_\Lambda \int_{-1}^{1} 
 d\cos\theta_{\ell} K(q^2,\theta_{\ell},\theta_\Lambda,\phi)} \label{Y3sc}\\
     &=& \frac{1}{2}\frac{K_{3sc}}{2K_{1ss}+K_{1cc}} \nonumber 
\end{eqnarray}

\begin{eqnarray}
 {Y_{4s}} &=&  \frac{\left[\int_{0}^{\pi} -\int_{\pi}^{2\pi}\right] d\phi \int_{-1}^{1} d\cos\theta_\Lambda \int_{-1}^{1} 
 d\cos\theta_{\ell} K(q^2,\theta_{\ell},\theta_\Lambda,\phi)}
 {\int_{0}^{2\pi} d\phi \int_{-1}^1 d\cos\theta_\Lambda \int_{-1}^{1} d\cos\theta_{\ell} K(q^2,\theta_{\ell},\theta_\Lambda,\phi)} \label{Y4s}\\
     &=& \frac{3\pi}{8}\frac{K_{4s}}{2K_{1ss}+K_{1cc}} \nonumber
     \end{eqnarray}
 and
 \begin{eqnarray}
  {Y_{4sc}} &=&  \frac{\left[\int_{0}^{\pi} -\int_{\pi}^{2\pi}\right] d\phi \int_{-1}^{1} d\cos\theta_\Lambda 
  \left[\int_{0}^{1} -\int_{-1}^0\right] d\cos\theta_{\ell} K(q^2,\theta_{\ell},\theta_\Lambda,\phi)}
 {\int_{0}^{2\pi} d\phi \int_{-1}^1 d\cos\theta_\Lambda \int_{-1}^{1} d\cos\theta_{\ell} 
 K(q^2,\theta_{\ell},\theta_\Lambda,\phi)} \label{Y4sc}\\
     &=& \frac{1}{2}\frac{K_{4sc}}{2K_{1ss}+K_{1cc}} \nonumber 
 \end{eqnarray}

Clearly, Eq.(\ref{Y2})-Eq.(\ref{Y4sc}) along with the other equations above determine all the angular coefficients. Although in
principle true, in practice any such determination will be severely hampered by the poorly known transition form factors. In the 
baryonic case, the form factors are rather poorly known  when one compares the situation with the mesonic
counterparts. In the latter, there has been lot of progress in having reliable set of form factors. But even there,
hadronic uncertainties prevent from making any sound claim of new physics when encountering deviations
from SM.

The kinematic region can be divided into the large and small $q^2$ or equivalently the low and large recoil regions.
In each of the regions, one can make suitable approximations which allow a smaller set of form factors to be employed, and
there are certain relations that emerge between various form factors. A typical matrix element one is interested in
is of the form: $\langle\Lambda(k,s_{\Lambda}\vert\bar{s}\Gamma b\vert\Lambda_b(p,s_{\Lambda_b}\rangle$,
where $s_{\Lambda_{(b)}}$ are the spin vectors associated with the baryons. In full
generality, there are a large number of form factors that would contribute to the physical decay rate. If, however, 
one makes use of the heavy quark symmetry (working systematically in heavy quark effective theory (HQET)), the number 
of independent form factors reduces to just two. Further, in the large recoil limit, there is only one
independent form factor. There exist several estimates of the form factors \cite{Mannel:1990vg}. However, only limited information is available
on form factors computed directly in QCD using lattice techniques  \cite{Detmold:2012vy}. Employing HQET, the two relevant form factors
appear in the hadronic matrix elements as:
\begin{equation}
 \langle\Lambda(k,s_{\Lambda}\vert\bar{s}\Gamma b\vert\Lambda_b(p,s_{\Lambda_b}\rangle = \bar{u}(k,s_{\Lambda})
 \left[F_1(k.v) + \not{p} F_2(k.v)\right]\Gamma \,{\mathcal{U}}(v,s_{\Lambda_b})
\end{equation}
where $v$ is the velocity of $\Lambda_b$ and the two form factors depend only on the invariant $k.v$, the energy
of $\Lambda$ in the rest frame of $\Lambda_b$. The spinors satisfy the relations
\begin{equation}
\sum_{s=1,2}u(p,s)\bar{u}(p,s) = m_{\Lambda}+\not{p}, \,\,\,\,\,\,\, \sum_{s=1,2}{\mathcal{U}}(v,s)\bar{{\mathcal{U}}}(v,s) = 1+\not{p}
\end{equation}
It turns out that the two linear combinations $F_{\pm} = F_1 \pm F_2$
are more useful and one therefore prefers to work with them. At present, the lattice calculations provide a reliable estimate of the form factors
$F_{\pm}$ (or $F_{1,2}$) only in the region $q^2\geq 13$ GeV$^2$. Below this $q^2$ value, only model dependent extrapolations are to be made and
relied on. But it is still reassuring to observe that the lattice results extrapolated over the whole $q^2$ range tend to give a reasonable fit
to the available experimental data. From the results of the lattice calculations \cite{Detmold:2012vy},
 one finds that over the whole kinematic range, the ratio of the two form factors
$F2$ and $F_1$ takes the value
\begin{equation}
-\frac{F_2}{F_1} \sim 0.21
\end{equation}
with a gentle variation as the baryon energy is varied: the ratio lies in the interval $[0.19,0.23]$. Further, using the relations between $F_{\pm}$ and $F_{1,2}$,
one finds that $F_+ \sim 0.8 F_1$ and $F_-\sim 1.2 F_1$. One therefore finds that in the heavy quark limit (strictly valid only when $m_b\to \infty$), 
there is essentially one form factor, especially in the large $q^2$ region. Incidentally, the recent LHCb measurements of the branching ratio
and the simplest angular asymmetries are also in the large $q^2$ region.

Although at the present level of accuracy, only one form factor seems to be enough (since due to
the lattice determination of the ratio $F_2/F_1$, which does not show much variation,
essentially one is dealing with only one form factor), 
it is clear that the other form factor will be needed to validate SM itself. To this end
it is worthwhile and important to construct observables which are as free of the hadronic inputs as possible. With this spirit we propose the following observables, all of which 
have a zero crossing point in the large $q^2$ region. This will allow to have a meaningful comparison with the experimental data as lattice results can be trusted in this
kinematic region and accurate numerical predictions can be obtained.
\begin{equation}
{\mathcal{T}}_1 = \frac{A^{R*}_{\perp 1}A^{R}_{\parallel 1} - A^{R*}_{\perp 0}A^{R}_{\parallel 0} - (R\to L)}{A^{R*}_{\perp 1}A^{R}_{\parallel 1} + A^{R*}_{\perp 0}A^{R}_{\parallel 0} +(R\to L)}
\end{equation}

\begin{equation}
{\mathcal{T}}_2 = \frac{A^{R*}_{\perp 1}A^{R}_{\parallel 1} + A^{R*}_{\perp 0}A^{R}_{\parallel 0} - 
(A^{L*}_{\perp 1}A^{L}_{\parallel 1} - A^{L*}_{\perp 0}A^{L}_{\parallel 0})}{A^{R*}_{\perp 1}A^{R}_{\parallel 1} + A^{R*}_{\perp 0}A^{R}_{\parallel 0} +
(A^{L*}_{\perp 1}A^{L}_{\parallel 1} + A^{L*}_{\perp 0}A^{L}_{\parallel 0})}
\end{equation}

and 
\begin{equation}
{\mathcal{T}}_3 = \frac{\vert A^{R}_{\perp 1}\vert^2 -  \vert A^{R}_{\perp 0}\vert^2 + (R\to L)}{\vert A^{R}_{\perp 1}\vert^2 +  \vert A^{R}_{\perp 0}\vert^2 + (R\to L)}
\end{equation}

The zero crossing points for these observables, particularly ${\mathcal{T}}_{1,2}$ are completely free of the form factors. They read (neglecting the small
imaginary part in the Wilson coefficient $C_9$, i.e., assuming it to be real for illustration):
\begin{eqnarray}
s_0({\mathcal{T}}_1) &=& \frac{M_{\Lambda_b}(C_9M_{\Lambda_b}F_+F_- - 2C_7m_bF_+F_-)}{2C_9F_+F_-} = \frac{M_{\Lambda_b}(C_9M_{\Lambda_b} - 2C_7m_b)}{2C_9}\\
&=& \frac{M_{\Lambda_b}^2}{2} - \frac{M_{\Lambda_b}m_bC_7}{C_9} \simeq \frac{M_{\Lambda_b}^2}{2} + \frac{s_0^{\ell}}{2} \nonumber\\
&\sim&17.52\,\, \mathrm{GeV}^2 \nonumber
\end{eqnarray}
where in the penultimate step we used the approximate relation of the leptonic forward-backward zero crossing. However, it should be immediately clear that
even without making use of this relation, $s_0({\mathcal{T}}_1)$ is a genuinely short distance quantity and therefore has a precise value within SM that can be
unambiguously compared with the experimental determination. Next consider the zero crossing point of  ${\mathcal{T}}_2 $:
\begin{equation}
s_0({\mathcal{T}}_2) = -\frac{M_{\Lambda_b}^2C_{10}^2 + 8M_{\Lambda_b}m_bC_{10}C_7 +(M_{\Lambda_b}C_9+2C_7m_b)^2}{4C_9C_{10}} \sim 17.63\,\, \mathrm{GeV}^2
\end{equation}
This is again a pure short distance quantity and like $s_0({\mathcal{T}}_1)$ turns out to be very clean. 
The zero crossing point of the third observable works out to be:
\begin{eqnarray}
 s_0({\mathcal{T}}_3) &=& \frac{1}{4(C_{10}^2+C_9^2)}\big[M_{\Lambda_b}^2(C_{10}^2+C_9^2) - 4M_{\Lambda_b}m_bC_{9}C_7 + 4 m_b^2C_7^2 \nonumber \\
 &+& \sqrt{(M_{\Lambda_b}^2C_{10}^2 + (M_{\Lambda_b}C_9-2C_7m_b^2))^2 - 64C_7^2M_{\Lambda_b}^2m_b^2(C_{10}^2+C_9^2)}\,\,\big] \\
 &\sim& 17.24\,\, \mathrm{GeV}^2 \nonumber
\end{eqnarray}
In the last observable, the replacement $\perp\to \parallel$ generates another observable which has the same zero crossing
point. One could simply combine these two into linear combinations and study the profiles.
It is quite evident that all the three zero crossings lie well in the large $q^2$ region where at present there is more control
theoretically. Recent LHCb results on the angular analysis of $B\to K^*\ell^+\ell^-$ have shown
deviations from SM expectations, especially for the observable $P_5'$. Many possible solutions have
been suggested, among which the minimal solution that gives a reasonably good fit is the solution
where the SM operator basis is employed and the only deviation is in $C_9$: $\delta C_9 \sim -1.5$, while
there are practically no deviations in the other two Wilson coefficients \cite{Descotes-Genon:2013wba}. Assuming this scenario,
it is clear that the above observables, in particular the zero crossings can, very effectively and
in a robust manner, test this hypothesis. In fact, extension to an extended operator basis is straight forward.
We thus immediately see the immense potential of these asymmetries and zero crossing points which are
very clean. The other advantage of these zero crossing points lies in the fact that they lie in
the high $q^2$ region, in sharp contrast to the zero crossings of the observables in $B\to K^*\ell^+\ell^-$.
This additional feature will also help in understanding possible $q^2$ dependence and differentiate
between possibly overlooked hadronic effect from genuine new physics contribution which by definition
should be $q^2$ independent. 

These zero crossing points, along with the zeroes of the leptonic and hadronic forward-backward asymmetries
can be simultaneously used to not only test SM but also to infer more about the form factors. Without making any assumptions,
these quantities are functions of various form factors (actually ratios of various form factors). Measurement of these
quantities, along with the profiles of various observables will allow us to extract some of these ratios at specific points.
This information can then be utilized to cross-check the consistency of the form factors that one has employed. At present, this
may appear as a daunting task but with more data available and more observables measured precisely, a simultaneous fit
will provide this valuable information. We hasten to add that there are several other angular observables that can be constructed
like:
\begin{equation}
 {\mathcal{O}}_1 = \frac{\vert A^{R}_{\perp 1}\vert^2 + \vert A^{R}_{\perp 0}\vert^2 - (R\to L)}
 {\vert A^{R}_{\perp 1}\vert^2 +  \vert A^{R}_{\perp 0}\vert^2 + (R\to L)}
\end{equation}
and 
\begin{equation}
 {\mathcal{O}}_2 = \frac{\vert A^{R}_{\parallel 1}\vert^2 + \vert A^{R}_{\parallel 0}\vert^2 - (R\to L)}
 {\vert A^{R}_{\parallel 1}\vert^2 +  \vert A^{R}_{\parallel 0}\vert^2 + (R\to L)}
\end{equation}
These are also interesting and measurable quantities and have zero crossing points, which lie in the low
$q^2$ region. Since at present both the form factor availability and experimental results are in the high
$q^2$ region, we do not pursue them further. A detailed numerical study, taking into account
all the asymmetries and observables discussed above will be presented elsewhere. One can systematically
study the effects of other operators and the impact they have on various observables, 
in particular zero crossings.

All what has been discussed so far has been in purview of form factors obtained with heavy quark limit
in mind. An immediate improvement would be to start with the HQET relations and try to include the first
sub-leading terms. In the absence of a lattice calculation along these lines, such an endeavour would be
phenomenological to some extent but would still be worthwhile. 

\section{Discussion and conclusions}
The standard model of particle physics has stood various experimental tests and there has not been
any evidence of new physics till now. However, the fact that experiments have unambiguously
established that neutrinos have mass, that there is dark matter in the universe and the cosmological
observations clearly showing that the energy budget of the universe is dominated by dark energy already
point to the fact that there is definitely something beyond SM. Further, the observed baryon
asymmetry of the universe can not be explained within SM, which falls short by orders of magnitude.
All these experimental evidences call for some extension of SM. Though not conclusive at the moment,
there are several anomalies in the flavour sector. Most recent ones are related to $b\to s$ semi-leptonic
decay modes. In this vain, it is important to study different modes and channels which are mediated by 
the same $b\to s$ fundamental interactions. Recent times have seen a lot of theoretical
and experimental effort in exploiting $B\to K^{(*)}\ell^+\ell^-$ modes to their full potential.
The corresponding baryonic mode $\Lambda_b\rightarrow \Lambda \ell^+\ell^-$ has started to be
studied experimentally. Since the baryonic counterpart now involves a completely
different set of hadronic inputs, this becomes a very useful playground to cross-check the
anomalies seen in the mesonic channels. Only recently, a more systematic approach to fully 
exploit the host of angular observables this mode has to offer has been initiated. In the present
work we have extended that effort and listed all the angular asymmetries that pin down
the complete set of angular coefficients. We have also proposed several other angular observables
which should be easy to access experimentally. At present, this theoretical effort is limited by our
knowledge of the baryonic form factors. The observables suggested in the present work are not
affected by this limitation and in fact are best suited to work in the region of validity of
the available form factors. These new observables are theoretically clean and therefore probe the genuine short
distance content of the underlying theory. Moreover, they have a zero crossing point in the large $q^2$ 
region, the region where reliable form factors are available from lattice calculations. Therefore,
this baryonic decay has an immense potential to test SM precisely and even with limited
amount of data available in near future, there may be a hope to have a good indication
of any new physics, if it is really there at the TeV scale.

Before closing, we would like to mention some of the possible improvements and future directions.
As is obvious, theoretical calculations are presently limited by the form factors. The available
form factors from lattice have been obtained in the strict heavy quark limit. One may attempt
to improve those by trying to include sub-leading terms, even if in somewhat approximate
manner. The other direction is to include other operators beyond SM and study the proposed
observables within the extended operator basis. This would shed some light on the (ir)relevance
of some operators. When combined with similar studies on the mesonic counterparts, this could
limit the beyond SM contributions significantly. In particular, a detailed numerical investigation
of the baryonic mode with inputs and recent hints of possible new physics from $B\to K^{(*)}\ell^+\ell^-$
would be very useful.

%\section{Angular Observables of $B\rightarrow K^* l^+l^-$ in large recoil limit}

%\bibliography{/home/girish/Desktop/citation}{}

\begin{thebibliography}{99}
%%GR
% The bibliography was re-sorted...
%%
% PRD: Phys. Rev. D {\bf nn}, ppppp (yyyy)
% PLB: Phys. Lett. B {\bf nn}, ppp (yyyy)
% PRL: Phys. Rev. Lett. {\bf nn}, ppp (yyyy)
% ZPC: Z. Phys. C {\bf nn}, ppp (yyyy)
% JHEP: J. High Energy Phys. 01, ppp (yyyy)
% EPJ: Eur. Phys. J. C {\bf nn}, ppp (yyyy)
% NPB: Nucl. Phys. {\bf Bnnn}, ppp (yyyy)
% {\it et al.}

\bibitem{Buchalla:1995vs} 
  G.~Buchalla, A.~J.~Buras and M.~E.~Lautenbacher,
  %``Weak decays beyond leading logarithms,''
  Rev.\ Mod.\ Phys.\  {\bf 68}, 1125 (1996)
  [hep-ph/9512380].
  %%CITATION = HEP-PH/9512380;%%


\bibitem{Aaij:2013qta} 
  R.~Aaij {\it et al.}  [LHCb Collaboration],
  %``Measurement of Form-Factor-Independent Observables in the Decay $B^{0} \to K^{*0} \mu^+ \mu^-$,''
  Phys.\ Rev.\ Lett.\  {\bf 111}, no. 19, 191801 (2013)
  [arXiv:1308.1707 [hep-ex]].
  %%CITATION = ARXIV:1308.1707;%%

\bibitem{Aaij:2014ora} 
  R.~Aaij {\it et al.} [LHCb Collaboration],
  %``Test of lepton universality using $B^{+}\rightarrow K^{+}\ell^{+}\ell^{-}$ decays,''
  Phys.\ Rev.\ Lett.\  {\bf 113}, 151601 (2014)
  [arXiv:1406.6482 [hep-ex]].
  %%CITATION = ARXIV:1406.6482;%%
  
  
  
  \bibitem{Huang:1998ek} 
  C.~S.~Huang and H.~G.~Yan,
  %``Exclusive rare decays of heavy baryons to light baryons: Lambda(b) ---> Lambda gamma and Lambda(b) ---> Lambda l+ l-,''
  Phys.\ Rev.\ D {\bf 59}, 114022 (1999)
  [Phys.\ Rev.\ D {\bf 61}, 039901 (2000)]
  [hep-ph/9811303];
  %%CITATION = HEP-PH/9811303;%%
  %\bibitem{Aliev:1999ap} 
  T.~M.~Aliev and M.~Savci,
  %``Exclusive Lambda(b) ---> Lambda lepton+ lepton- decay in two Higgs doublet model,''
  J.\ Phys.\ G {\bf 26}, 997 (2000)
  [hep-ph/9906473];
  %%CITATION = HEP-PH/9906473;%%
  %\bibitem{Chen:2001ki} 
  C.~H.~Chen and C.~Q.~Geng,
  %``Rare Lambda(b) ---> Lambda lepton+ lepton- decays with polarized lambda,''
  Phys.\ Rev.\ D {\bf 63}, 114024 (2001)
  [hep-ph/0101171];
  %%CITATION = HEP-PH/0101171;%%
  %\bibitem{Chen:2001sj} 
  C.~H.~Chen and C.~Q.~Geng,
  %``Lepton asymmetries in heavy baryon decays of Lambda(b) ---> Lambda lepton+ lepton-,''
  Phys.\ Lett.\ B {\bf 516}, 327 (2001)
  [hep-ph/0101201];
  %%CITATION = HEP-PH/0101201;%%
  %\bibitem{Aliev:2002hj} 
  T.~M.~Aliev, A.~Ozpineci and M.~Savci,
  %``New physics effects in $\Lambda_b \to \Lambda \ell^{+} \ell^{-}$ decay with lepton polarizations,''
  Phys.\ Rev.\ D {\bf 65}, 115002 (2002)
  [hep-ph/0203045];
  %%CITATION = HEP-PH/0203045;%%
  %\bibitem{Aliev:2002nv} 
  T.~M.~Aliev, A.~Ozpineci, M.~Savci and C.~Yuce,
  %``$T$ violation in $\Lambda_b \to \Lambda \ell^{+} \ell^{-}$ decay beyond standard model,''
  Phys.\ Lett.\ B {\bf 542}, 229 (2002)
  [hep-ph/0206014];
  %%CITATION = HEP-PH/0206014;%%
  %\bibitem{Aliev:2002tr} 
  T.~M.~Aliev, A.~Ozpineci and M.~Savci,
  %``Model independent analysis of Lambda baryon polarizations in Lambda(b) ---> Lambda l+ l- decay,''
  Phys.\ Rev.\ D {\bf 67}, 035007 (2003)
  [hep-ph/0211447];
  %%CITATION = HEP-PH/0211447;%%
  %\bibitem{Aliev:2004af} 
  T.~M.~Aliev, V.~Bashiry and M.~Savci,
  %``Forward-backward asymmetries in Lambda(b) ---> Lambda l+ l- decay beyond the standard model,''
  Nucl.\ Phys.\ B {\bf 709}, 115 (2005)
  [hep-ph/0407217];
  %%CITATION = HEP-PH/0407217;%%
  %\bibitem{Aliev:2004yf} 
  T.~M.~Aliev, V.~Bashiry and M.~Savci,
  %``Double-lepton polarization asymmetries in Lambda(b) ---> Lambda l+ l- decay,''
  Eur.\ Phys.\ J.\ C {\bf 38}, 283 (2004)
  [hep-ph/0409275];
  %%CITATION = HEP-PH/0409275;%%
 %\bibitem{Giri:2005yt} 
  A.~K.~Giri and R.~Mohanta,
  %``Effect of R-parity violation on the rare decay Lambda/b ---> Lambda mu+ mu-,''
  J.\ Phys.\ G {\bf 31}, 1559 (2005);
  %%CITATION = JPAGA,G31,1559;%% 
  %\bibitem{Aliev:2006xd} 
  T.~M.~Aliev and M.~Savci,
  %``Lambda(b) ---> Lambda l+ l- decay in universal extra dimensions,''
  Eur.\ Phys.\ J.\ C {\bf 50}, 91 (2007)
  [hep-ph/0606225];
  %%CITATION = HEP-PH/0606225;%%
  %\bibitem{Aliev:2006gv} 
  T.~M.~Aliev, M.~Savci and B.~B.~Sirvanli,
  %``Double-lepton polarization asymmetries in Lambda(b) ---> Lambda l+ l- decay in universal extra dimension model,''
  Eur.\ Phys.\ J.\ C {\bf 52}, 375 (2007)
  [hep-ph/0608143];
  %%CITATION = HEP-PH/0608143;%%
  %\bibitem{Zolfagharpour:2007eh} 
  F.~Zolfagharpour and V.~Bashiry,
  %``Double Lepton Polarization in Lambda(b) ---> Lambda l+ l- Decay in the Standard Model with Fourth Generations Scenario,''
  Nucl.\ Phys.\ B {\bf 796}, 294 (2008)
  [arXiv:0707.4337 [hep-ph]];
  %%CITATION = ARXIV:0707.4337;%%
  %\bibitem{Wang:2008sm} 
  Y.~m.~Wang, Y.~Li and C.~D.~Lu,
  %``Rare Decays of Lambda(b) ---> Lambda + gamma and Lambda(b) ---> Lambda + l+ l- in the Light-cone Sum Rules,''
  Eur.\ Phys.\ J.\ C {\bf 59}, 861 (2009)
  [arXiv:0804.0648 [hep-ph]];
  %%CITATION = ARXIV:0804.0648;%%
  %\bibitem{Aslam:2008hp} 
  M.~J.~Aslam, Y.~M.~Wang and C.~D.~Lu,
  %``Exclusive semileptonic decays of Lambda(b) ---> Lambda l+ l- in supersymmetric theories,''
  Phys.\ Rev.\ D {\bf 78}, 114032 (2008)
  [arXiv:0808.2113 [hep-ph]];
  %%CITATION = ARXIV:0808.2113;%%
  %\bibitem{Aliev:2010uy} 
  T.~M.~Aliev, K.~Azizi and M.~Savci,
  %``Analysis of the $Lambda_{b}\rightarrow \Lambda \ell^+\ell^- $ decay in QCD,''
  Phys.\ Rev.\ D {\bf 81}, 056006 (2010)
  [arXiv:1001.0227 [hep-ph]];
  %%CITATION = ARXIV:1001.0227;%%
  %\bibitem{Sahoo:2009zz} 
  S.~Sahoo, C.~K.~Das and L.~Maharana,
  %``Effect of both Z and Z-prime mediated flavor-changing neutral currents on the baryonic rare decay Lambda(b) ---> Lambda l+ l-,''
  Int.\ J.\ Mod.\ Phys.\ A {\bf 24}, 6223 (2009)
  [arXiv:1112.4563 [hep-ph]];
  %%CITATION = ARXIV:1112.4563;%%
  %\bibitem{Mott:2011cx} 
  L.~Mott and W.~Roberts,
  %``Rare dileptonic decays of $\Lambda_b$ in a quark model,''
  Int.\ J.\ Mod.\ Phys.\ A {\bf 27}, 1250016 (2012)
  [arXiv:1108.6129 [nucl-th]];
  %%CITATION = ARXIV:1108.6129;%%
  %\bibitem{Gutsche:2013pp} 
  T.~Gutsche, M.~A.~Ivanov, J.~G.~Korner, V.~E.~Lyubovitskij and P.~Santorelli,
  %``Rare baryon decays $\Lambda_b \to \Lambda {l^{+}l^{-}} (l=e, \mu, \tau)$ and $\Lambda_b \to \Lambda\gamma$ : differential and total rates, lepton- and hadron-side forward-backward asymmetries,''
  Phys.\ Rev.\ D {\bf 87}, 074031 (2013)
  [arXiv:1301.3737 [hep-ph]].
  %%CITATION = ARXIV:1301.3737;%%
   
  
  
  
  \bibitem{Boer:2014kda} 
  P.~B\"{o}er, T.~Feldmann and D.~van Dyk,
  %``Angular Analysis of the Decay $\Lambda_b \to \Lambda (\to N \pi) \ell^+\ell^-$,''
  JHEP {\bf 1501}, 155 (2015)
  [arXiv:1410.2115 [hep-ph]].
  %%CITATION = ARXIV:1410.2115;%%

  
  
  \bibitem{Aaltonen:2011qs} 
  T.~Aaltonen {\it et al.} [CDF Collaboration],
  %``Observation of the Baryonic Flavor-Changing Neutral Current Decay $\Lambda_{b} \to \Lambda \mu^{+} \mu^{-}$,''
  Phys.\ Rev.\ Lett.\  {\bf 107}, 201802 (2011)
  [arXiv:1107.3753 [hep-ex]].
  %%CITATION = ARXIV:1107.3753;%%
  
  
  \bibitem{Aaij:2013mna} 
  R.~Aaij {\it et al.} [LHCb Collaboration],
  %``Measurement of the differential branching fraction of the decay $\Lambda_b^0\rightarrow\Lambda\mu^+\mu^-$,''
  Phys.\ Lett.\ B {\bf 725}, 25 (2013)
  [arXiv:1306.2577 [hep-ex]].
  %%CITATION = ARXIV:1306.2577;%%
  
  
  \bibitem{Aaij:2015xza} 
  R.~Aaij {\it et al.} [LHCb Collaboration],
  %``Differential branching fraction and angular analysis of $\Lambda^{0}_{b} \rightarrow \Lambda \mu^+\mu^-$ decays,''
  JHEP {\bf 1506}, 115 (2015)
  [arXiv:1503.07138 [hep-ex]].
  %%CITATION = ARXIV:1503.07138;%%
  
  
  
  
     
  
  \bibitem{Detmold:2012vy} 
  W.~Detmold, C.-J.~D.~Lin, S.~Meinel and M.~Wingate,
  %``?b???+?- form factors and differential branching fraction from lattice QCD,''
  Phys.\ Rev.\ D {\bf 87}, 074502 (2013)
  [arXiv:1212.4827 [hep-lat]].
  %%CITATION = ARXIV:1212.4827;%%
  
  
  
  \bibitem{Mannel:1990vg} 
  T.~Mannel, W.~Roberts and Z.~Ryzak,
  %``Baryons in the heavy quark effective theory,''
  Nucl.\ Phys.\ B {\bf 355}, 38 (1991);
  %%CITATION = NUPHA,B355,38;%%
  %\bibitem{Hussain:1990uu} 
  F.~Hussain, J.~G.~Korner, M.~Kramer and G.~Thompson,
  %``On heavy baryon decay form-factors,''
  Z.\ Phys.\ C {\bf 51}, 321 (1991);
  %%CITATION = ZEPYA,C51,321;%%
  %\bibitem{Hussain:1992rb} 
  F.~Hussain, D.~S.~Liu, M.~Kramer, J.~G.~Korner and S.~Tawfiq,
  %``General analysis of weak decay form-factors in heavy to heavy and heavy to light baryon transitions,''
  Nucl.\ Phys.\ B {\bf 370}, 259 (1992);
  %%CITATION = NUPHA,B370,259;%%
  %\bibitem{Cheng:1995fe} 
  H.~Y.~Cheng and B.~Tseng,
  %``1/M corrections to baryonic form-factors in the quark model,''
  Phys.\ Rev.\ D {\bf 53}, 1457 (1996)
  [Phys.\ Rev.\ D {\bf 55}, 1697 (1997)]
  [hep-ph/9502391];
  %%CITATION = HEP-PH/9502391;%%
  %\bibitem{Wang:2009hra} 
  Y.~M.~Wang, Y.~L.~Shen and C.~D.~Lu,
  %``Lambda(b) ---> p, Lambda transition form factors from QCD light-cone sum rules,''
  Phys.\ Rev.\ D {\bf 80}, 074012 (2009)
  [arXiv:0907.4008 [hep-ph]];
  %%CITATION = ARXIV:0907.4008;%%
%\bibitem{Mannel:2011xg} 
  T.~Mannel and Y.~M.~Wang,
  %``Heavy-to-light baryonic form factors at large recoil,''
  JHEP {\bf 1112}, 067 (2011)
  [arXiv:1111.1849 [hep-ph]];
  %%CITATION = ARXIV:1111.1849;%%
  %\bibitem{Feldmann:2011xf} 
  T.~Feldmann and M.~W.~Y.~Yip,
  %``Form Factors for $Lambda_b \to \Lambda$ Transitions in {SCET},''
  Phys.\ Rev.\ D {\bf 85}, 014035 (2012)
  [Phys.\ Rev.\ D {\bf 86}, 079901 (2012)]
  [arXiv:1111.1844 [hep-ph]];
  %%CITATION = ARXIV:1111.1844;%%
  %\bibitem{Wang:2011uv} 
  W.~Wang,
  %``Factorization of Heavy-to-Light Baryonic Transitions in SCET,''
  Phys.\ Lett.\ B {\bf 708}, 119 (2014)
  [arXiv:1112.0237 [hep-ph]].
  %%CITATION = ARXIV:1112.0237;%%
  
  
  
  
  
  \bibitem{Descotes-Genon:2013wba} 
  S.~Descotes-Genon, J.~Matias and J.~Virto,
  %``Understanding the $B\to K^*\mu^+\mu^-$ Anomaly,''
  Phys.\ Rev.\ D {\bf 88}, 074002 (2013)
  [arXiv:1307.5683 [hep-ph]];
  %%CITATION = ARXIV:1307.5683;%%
%\bibitem{Altmannshofer:2013foa} 
  W.~Altmannshofer and D.~M.~Straub,
  %``New physics in $B \to K^*\mu\mu$?,''
  Eur.\ Phys.\ J.\ C {\bf 73}, 2646 (2013)
  [arXiv:1308.1501 [hep-ph]];
  %%CITATION = ARXIV:1308.1501;%%
  %\bibitem{Gauld:2013qba} 
  R.~Gauld, F.~Goertz and U.~Haisch,
  %``On minimal $Z'$ explanations of the $B\to K^*\mu^+\mu^-$ anomaly,''
  Phys.\ Rev.\ D {\bf 89}, 015005 (2014)
  [arXiv:1308.1959 [hep-ph]];
  %%CITATION = ARXIV:1308.1959;%%
  %\bibitem{Descotes-Genon:2015uva} 
  S.~Descotes-Genon, L.~Hofer, J.~Matias and J.~Virto,
  %``Global analysis of $b\to s\ell\ell$ anomalies,''
  arXiv:1510.04239 [hep-ph].
  %%CITATION = ARXIV:1510.04239;%%
  
  
  
  
  

\end{thebibliography}
%\bibliographystyle{apsrev4-1}

%

\end{document}